\documentclass[10pt]{article}
\usepackage{fancyhdr}
\usepackage{extramarks}
\usepackage{amsmath}
\usepackage{amsthm}
\usepackage{amsfonts}
\usepackage{siunitx}
\usepackage{tikz}
\usepackage[plain]{algorithm}
\usepackage{algpseudocode}
\usepackage{multirow}
\usepackage{booktabs}
\usepackage{graphicx}
\usepackage{subfigure}
\usepackage[colorlinks,linkcolor=black,anchorcolor=black,citecolor=black,urlcolor=blue]{hyperref}
\usepackage{amsmath,bm}
\usepackage{booktabs}
\usepackage{mathtools}
\usepackage{amssymb}
\usepackage{caption}
\usepackage{capt-of}
\usepackage{mciteplus}
\usepackage{cite}
\usepackage{mathrsfs}
\usepackage[title,titletoc,toc]{appendix}
\usepackage{xr}
\usepackage{parskip}
\usepackage{soul}
\usepackage{textcomp}
\usepackage[colaction]{multicol}
\usepackage[switch]{lineno}
\usepackage{lipsum}
\usepackage{etoolbox}
\usepackage{longtable}
\usepackage{array}
\usepackage{tablefootnote}
\usepackage{ragged2e}
\newcolumntype{C}[1]{>{\centering\arraybackslash}p{#1}}
\usetikzlibrary{automata,positioning}
\usetikzlibrary{automata,positioning}

\begin{document}

	\title{Machine-learning Analysis of Opioid Use Disorder Informed by MOR, DOR, KOR, NOR and ZOR-Based Interactome Networks
	}
	
	\author{ Hongsong Feng$^1$, Rana Elladki$^1$, Jian Jiang$^4$, and Guo-Wei Wei$^{1,2,3}$\footnote{
			Corresponding author.		Email: weig@msu.edu} \\
		$^1$ Department of Mathematics, \\
		Michigan State University, MI 48824, USA.\\
		Michigan State University, MI 48824, USA.\\
		$^2$Department of Electrical and Computer Engineering,\\
		Michigan State University, MI 48824, USA. \\
		$^3$ Department of Biochemistry and Molecular Biology,\\
		Michigan State University, MI 48824, USA. \\
		$^4$ Research Center of Nonlinear Science, School of Mathematical and Physical Sciences,\\
		Wuhan Textile University, Wuhan, 430200, P R. China\\
		
	}

	\date{\today} 
	
	\maketitle

Opioid use disorder (OUD)  continuously poses major public health challenges and social implications worldwide with dramatic rise of opioid dependence leading to potential abuse. Despite that a few pharmacological agents have been approved for OUD treatment, the efficacy of said agents for OUD requires further improvement in order to provide safer and more effective pharmacological and psychosocial treatments. Preferable therapeutic treatments of OUD rely on the advances in understanding the neurobiological mechanism of opioid dependence. Proteins including mu, delta, kappa, nociceptin, and zeta opioid receptors are the direct targets of opioids. Each receptor has a large protein-protein interaction (PPI) network, that behaves differently when subjected to various treatments, thus increasing the complexity in the drug development process for an effective opioid addiction treatment. The report below analyzes the work by presenting a PPI-network informed machine-learning study of OUD. We have examined more than 500 proteins in the five opioid receptor networks and subsequently collected 74 inhibitor datasets. Machine learning models were constructed by pairing gradient boosting decision tree (GBDT) algorithm with two advanced natural language processing (NLP)-based molecular fingerprints. With these models, we systematically carried out evaluations of screening and repurposing potential of drug candidates for four opioid receptors. In addition, absorption, distribution, metabolism, excretion, and toxicity (ADMET) properties were also considered in the screening of potential drug candidates. Our study can be a valuable and promising tool of pharmacological development for OUD treatments.

	
	\pagenumbering{roman}
	\begin{verbatim}
	\end{verbatim}

	\newpage
	\clearpage
	\pagebreak
	{\setcounter{tocdepth}{4} \tableofcontents}
	\newpage
	
	\setcounter{page}{1}
	\renewcommand{\thepage}{{\arabic{page}}}
	
		\section{Introduction}

	Over three million people in the United States are currently suffering or have previously suffered from opioid use disorder (OUD). In 2020 alone, over 68,000 deaths were recorded from an overdose. Unfortunately, the numbers are continuously rising, and have more than tripled throughout the past 10 years. The opioid crisis or epidemic presents a substantial public health concern and costs the United States billions of dollars annually. It takes many components, including increased public awareness, improved economic conditions, and better therapies, to fully address the OUD crisis.  Since the treatment of OUD with medications is effective in reducing symptoms of drug withdrawal and cravings \cite{veilleux2010review}, there is a pressing need to further search and develop more effective OUD treatments.
		
		
		Opioid is a broad term for any natural or synthetic substance that binds to specific opioid receptors in the human body. It is widely used as analgesics medications in modern pain management. The three main brain receptors that opioids bind to are the mu opioid receptor (MOR), kappa opioid receptor (KOR), and delta-opioid receptor (DOR) in the central nervous system (CNS) and peripheral organs \cite{zaki1996opioid}, which are responsible for a plethora of physiological functions, such as analgesia, respiration, and hormonal regulation. Synthetic and exogenous opioids (i.e., morphine, heroin, oxycontin, etc.) act on the opioid receptors as endorphins, and repeated exposure to escalating use of opioids causes gradual adaptations in the brain \cite{veilleux2010review}. Tolerance develops and leads to heightened uncontrolled intake of drugs. Consequently, physical dependence emerges with drug craving to reduce withdrawal symptoms upon abstinence \cite{kosten2002neurobiology}. When opioids are improperly ingested, the interaction with the opioid receptors can induce a harmful effect on the CNS impacting the respiratory system and causing irreversible brain damage \cite{wang2019historical}. According to the Centers for Disease Control and Prevention (CDC), methadone, oxycodone, and hydrocodone contribute to the highest number of fatalities, resulting from opioid overdose. 
		
	 	MOR is critical in brain reward circuits. It has an important role in goal-directed behavior such as drug-seeking behavior and represents a major factor in the initiation of addictive behaviors. In the development of opioid addiction, poor decision-making, and cognition impairment, MOR translates the goal-directed behaviors to habitual behaviors, promoting compulsive drug use \cite{wang2019opioid}. Through experiments ran on animals, with similar physiological functions, the results demonstrated that MOR is pivotal in mediating therapeutic and adverse activities of opioids and is associated with the maintenance of drug use, drug craving, and relapse \cite{gerrits2003drug}. KOR has anti-reward effects and can induce dysphoria. Upon long-term exposure to opioids, KOR has impact on modifying the brain’s reward circuits, leading to a relapse \cite{bruchas2010dynorphin}. The activation of KORs suppresses unpleasant MOR/DOR-mediated side effects including the rewarding effect. KOR blockade may beneficially alleviate stress responses, reduce drug cravings, and remediate depressive states. DORs reduce levels of anxiety and attenuate depressive symptoms \cite{roberts2001increased}. In addition, beneficial effects of DOR agonists were found in treating chronic pain and psychiatric disorders \cite{pradhan2011delta}.

		In growing efforts to combat the opioid epidemic, further research and development have been invested in the treatments of OUD. Currently, there are three medications used to treat opioid dependency approved by the Food and Drug Administration (FDA), methadone, buprenorphine, and naltrexone. Methadone is a full agonist on MOR and is used to reduce withdrawal and craving symptoms in patients. Methadone maintenance treatment (MMT) is useful in reducing the intensity of withdrawal symptoms and preventing patients from ingesting more opioids to induce a euphoric effect. Since MMT can decrease the intensity of cravings in patients, they are more willing to remain in treatment. However, methadone is associated with the risk of causing respiratory depression when improperly administrated \cite{modesto2010methadone}. Buprenorphine, a partial MOR agonist, is an alternative to methadone. It has a ceiling effect of stimulation on MOR than that of methadone. Hence, buprenorphine provides a less euphoric effect and is less likely to cause respiratory depression. MMT is more likely to keep patients in treatment than buprenorphine treatment \cite{mattick2014buprenorphine}. Naltrexone is an antagonist of MOR and is effective in attenuating drug cravings and reducing the risk of overdose. It does not produce sedation, analgesia, euphoria, or potential for abuse or diversion \cite{gastfriend2011intramuscular}. However,  it is not as widely used as methadone or buprenorphine for several reasons including low rates of patient acceptance and non-adherence. Naloxone is a non-selective and competitive opioid receptor antagonist and is used in treating opioid overdose or for opioid intoxication such as reversing respiratory depression. Take-home naloxone programs were developed to prevent fatal overdoses \cite{bell2020medication}. Buprenorphine/naloxone formulations are adopted. When injected, naloxone has higher bioavailability, thereby blocking the pain and craving-reducing effects of buprenorphine \cite{a2011buprenorphine}. However, Naloxone's capabilities are limited when ingesting highly potent opioids, such as fentanyl \cite{national2017naloxone}. Psychosocial interventions were also combined with these medications to improve the efficacy in treating opioid addictions \cite{gonzalez2004combating}. Further studies on medication efficacy and the mechanism of opioid addiction in the brain are needed to find better treatments to prevent relapse and facilitate longer periods of abstinence.

		The molecular mechanism underlying opioid tolerance, dependence, withdrawal, and addiction is complicated, involving several systems located in different regions of the brain. Opioids target opioid receptors in the brain and activate the mesolimbic (midbrain) reward system. Dopamine is produced in the ventral tegmental area (VTA), and is then released into the nucleus accumbens (NAc) area, giving rise to the feeling of pleasure. Opoid tolerance occurs because of the brain's adaptation to repeated exposure to opioids. \cite{veilleux2010review} The withdrawal symptoms and opioid dependence are related to noradrenaline (NA) that is produced in the locus ceruleus (LC) area \cite{kosten2002neurobiology}. Opioids impact brain areas with a fairly large number of proteins and peptides that are responsible for a multitude of physiological and biological functions. It is challenging to understand how so many proteins are simultaneously impacted by opioids, causing difficulty when designing effective medications for OUD treatments. On one hand, medication compounds targeting opioid receptors can potentially cause unintentional dependence or an overdose such as methadone, due to the possibility of an agonist effect on MOR. On the other hand, blocking other proteins associated with the opioids can interfere with the biological functions of these proteins and induces various side effects. It is necessary to investigate the inhibition effects of compounds on the opioid receptors as well as the side effects of potential medications by blocking other proteins. 
		
	The protein-protein interaction (PPI) network on the proteome scale forms a basis for systematically studying potential treatment efficacy and side effects. A PPI network is constituted of proteins and corresponding direct and indirect interactions that contribute to certain biological activities. The String v11 database (https://string-db.org/) \cite{szklarczyk2019string} provides a large collection of protein-protein interactions for given proteins or diseases. In the study of OUD, we can extract the PPI networks related to the major opioid receptors, based on which we can have systematic investigations of medication treatment and side effects. The proteins in these PPI networks are the test targets of treatment or side effects but using traditional in vivo or in vitro experiments is too time-consuming and expensive. Besides, large-scale experiments on animals raise legal and ethical concerns. Machine learning/deep learning technology has recently gained wide popularity in drug discovery and development, such as the generation of drug-like molecules \cite{gao2020generative}, repositioning of existing drugs for diseases \cite{gao2020repositioning}, protein engineering \cite{qiu2022persistent} and predictions of chemical toxicity \cite{yang2018silico} in drug design. The time and cost can be significantly reduced by machine learning as well as the erasure of ethical concerns. As a result, machine-learning approaches were utilized in this study, to carry out large-scale predictions. 
		
		In this work, we developed a proteome-informed machine-learning (ML) platform for the discovery of   anti-opioid addiction compounds. From the String v11 database, we obtained PPI networks of the five major opioid receptors with the associated proteins regarded as potential treatment and side effect targets. We then collected inhibitor datasets with experimental binding affinity labels from ChEMBL database for these protein targets and built machine learning models. The inhibitor compounds were represented by two forms of latent-vector (LV) fingerprints generated by transformer and autoencoder learning models, respectively. These latent vectors were paired with gradient boosting decision algorithm (GBDT) in building our binding affinity (BA) predictors. We then carried out cross-predictions to screen side effects and repurposing potentials of more than 120,000 compounds. With these models, we had more side effect evaluations of FDA-approved drugs or other existing medications. Another application using these cross-prediction models was to find promising lead compounds. In addition to the concern in potency and side effect, we also considered evaluations of pharmacokinetic properties in compound filtering , i.e., absorption, distribution, metabolism, excretion, and toxicity (ADMET) as well as synthesizability. Our platform is believed to be useful in advancing the drug development in treating OUD.

		\section{Results}

		\subsection{The Opioid receptors and addiction PPI networks.}
		
		\begin{figure}[ht]
			\centering
			\includegraphics[width=1.0\linewidth]{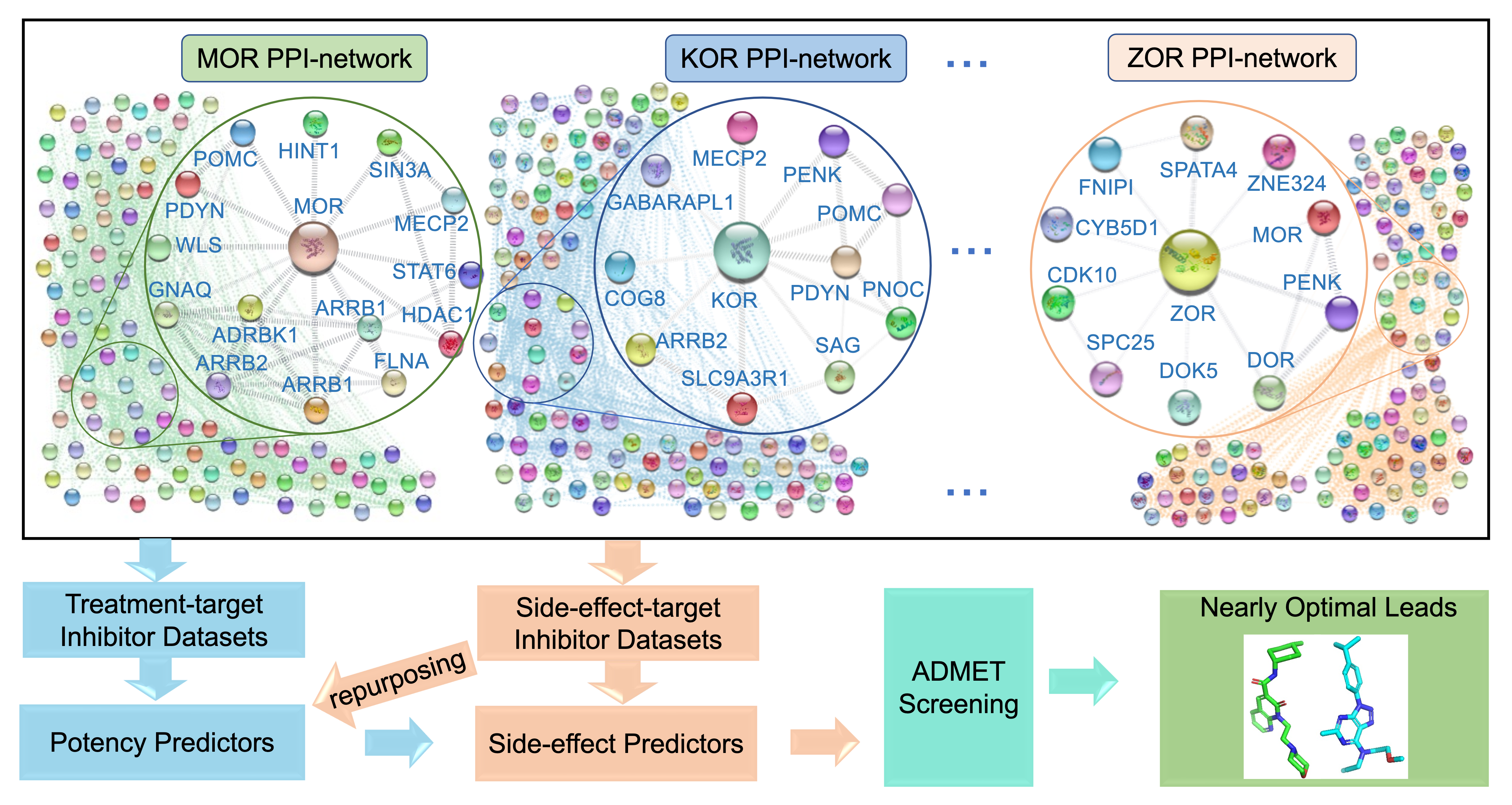} 
			\caption{{\footnotesize The workflow for searching nearly optimal lead compounds. The inhibitor compounds were collected for the proteins in protein-protein interaction networks of the five opioid receptors including mu, kappa, delta, nociceptin, and zeta opioid receptors. Each receptor has a core and global PPI network. The proteins in one core network have direct interaction with the opioid receptor, and those in a global network are related to opoid receptors through the proteins in the core network. Abbreviations for the proteins in the core networks are shown. Full names of the proteins in the five core networks are provided in the Supporting information.} }
			\label{Fig:workflow}
		\end{figure}
		
		Opioid receptors play critical roles in opioid dependence and are often the pharmacological targets of medications. There are four major subtypes of opioid receptors, namely mu opioid receptor (MOR), delta opioid receptor (DOR), kappa opioid receptor (KOR), and nociceptin opioid receptor (NOR). In addition, zeta opioid receptor (ZOR) is also believed to be an important one. However, ZOR was recently discovered, hence less studied and shares little sequence similarity with other opioid receptors. Opioid receptors are crucial in various biological functions and have broad distributions in the brain, spinal cord, on peripheral neurons, and digestive tract. MOR, KOR, and DOR are closely related to analgesia, opioid dependence and the adverse effect of respiratory depression caused by opioids, but each of them is distributed in various regions of the brain. NOR is distributed mainly in the cortex amygdala, hippocampus, septal nuclei, habenula, and hypothalamus in the brain and spinal cord, and is linked to development of tolerance to MOR agonists. ZORs widely exist in many parts of the body including the heart, liver, kidneys and brain. Its functions are mainly on tissue growth. Several clinically useful medications for treating addiction target MOR, KOR, and DOR \cite{veilleux2010review}, but the roles of NOR and ZOR in causing opioid dependence has not been much explored. However, they cannot be neglected in the pharmaceutical treatment of opioid dependence as they are all critical targets of opioids.
		
		  Opioid receptors have wide distributions in the body, and the synergistic interactions between these receptors and many other proteins upstream and downstream contribute to specific biological functions. As discussed before, we carry out drug discovery in the  PPI  networks. We extracted five PPI networks centered around each of the five opioid receptors by inputting receptor names, namely, mu-opioid receptor, delta-opioid receptor, kappa opioid receptor, zeta opioid receptor, nociceptin opioid receptor, and OGFR into the String database. In each network, there is a core subnetwork with proteins interacting directly with each opioid receptor, while proteins with direct and indirect interactions jointly form the global network as shown in Figure \ref{Fig:workflow}. We restrict the number of proteins in each global network to 101. Although more proteins should be considered, we limit our efforts to critical ones. There are five global networks in which five core networks exist. MOR, KOR, DOR, NOR, and ZOR are the most important proteins in the networks as each core protein plays an essential role. The five networks are not independent of each other with a few overlapping proteins found between the networks.
		
		Compounds with agonist or antagonist effects on opioid receptors showed their pharmacological effects in treating opioid dependence \cite{wang2019historical}, hence encouraging us to look for more compounds that bind to the opioid receptors. A desired drug must be specific to a target protein without causing adverse side effects to the other proteins. To evaluate the binding effect of inhibitors to receptor proteins and other proteins in the PPI networks, we collected inhibitor compounds from the ChEMBL database for each protein and built machine-learning models. We then used them to systematically analyze the side effects and repurposing potential of inhibitor compounds. We collected 74 datasets in total, with sufficient inhibitor data points for the proteins in the five extracted PPI networks with a total of 129,515 inhibitor compounds. In addition, we collected an inhibitor dataset for hERG protein and built an appropriate machine-learning model. The hERG is a critical potassium channel that must be avoided in drug design and discovery, as the blockade of the hERG channel is associated with prolongation of the long QT syndrome, eventually leading to fatal arrhythmia, namely Torsade de Pointes (TdP) \cite{sanguinetti2006herg}. In total, we collected 75 protein targets and built 75 machine-learning models. Since ZOR was recently discovered, not much inhibitor data was available to build a model for the ZOR protein. However, we were able to build models for the four remaining receptors, i.e., MOR, KOR, DOR, and NOR. Further, we used all the models to explore potential drugs that bind to the four opioid receptors. The details about the collected datasets can be found in the Supporting information.

		\subsection{Binding affinity predictions}
		
\begin{figure}[ht]
			\centering
			\includegraphics[width=0.85\linewidth]{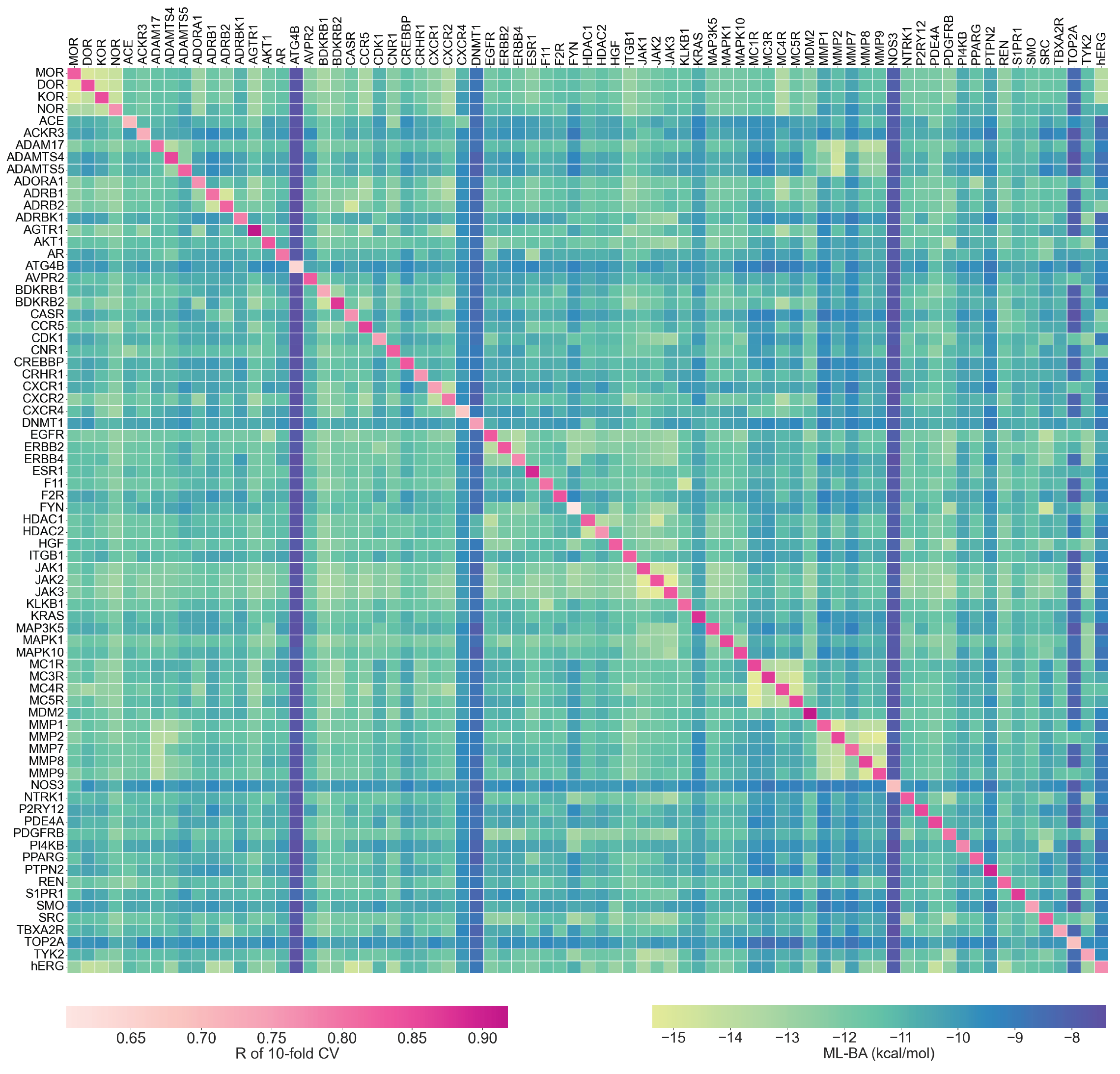} 
			\caption{{\footnotesize The heatmap of cross-target binding affinities (BAs) predictions revealing the inhibitor specificity of each dataset on other protein targets. The notations above the heatmap shows the machine-learning models while the those on the left of the heatmap denote all the inhibitor datasets. The diagonal elements in the heatmap indicate the Pearson correlation efficient (R) of five-fold cross validations for all the predictive models. The off-diagonal elements in each row represent the highest BAs values of inhibitors in one dataset predicted by 74 machine-learning models.} }
			\label{Fig:heatmap}
		\end{figure}

		The heatmap in Figure~\ref{Fig:heatmap} shows the cross-target binding affinity (BA) predictions using the 75 machine-learning models. The diagonal elements indicate the Pearson correlation coefficient (R) of five-fold cross-validation for our machine-learning models. Two of the 75 models have R values greater than 0.9, and the R values for fifty of them are greater than 0.8. The minimal R value of 0.604 is from the model built with the FYN inhibitor dataset. Overall, these models show excellent prediction accuracy and are reliable for BA predictions. 
		
		\subsubsection{Cross-target binding affinity predictions}
		
		Cross-target predictions can reveal side effects of drug candidates on other proteins. The off-diagonal elements represent the maximal BA values (BA with the largest absolute value) of inhibitor compounds in one dataset predicted by other models. The notations to the left of the heatmap indicate the 75 inhibitor datasets and those on top of the heatmap represent all the 75 machine-learning models. Each column exhibits all the predictions by one model. Specifically, the $i$-th element in the $j$-th column is the prediction result of  $i$-th dataset by the $j$-th model. These cross-target prediction results are indicators of side effects of one inhibitor dataset on other proteins. The BA value of -9.54 kcal/mol (K$_i$= 0.1 $\mu$M) is widely accepted as an inhibition threshold in the literature \cite{flower2002drug}. With this threshold, 5103 out of the 5625 cross-predictions were found to have side effects, i.e., the predicted maximal BA less than -9.54 kcal/mol. On the other hand, the remaining 522 cross-prediction results with maximal BA greater than -9.54 kcal/mol suggest  weak side effects. The color of the off-diagonal elements indicates the strength of side effects. The lighter the color, the stronger the side effects are. 
		
		Similar binding sites on off-target proteins is one of many reasons for side effects caused by drug candidates for one designated protein. Proteins in the same family can have similar three-dimensional (3D) structures or protein sequences, giving rise to the existence of similar binding sites. An inhibitor compound potent at one protein likely binds to another protein in the same family. As observed in Figure~\ref{Fig:heatmap}, mutual side effects occur among the four opioid receptors, i.e., MOR, DOR, KOR, and NOR. The 16 yellow square boxes on the upper left corner of the heatmap showed the cross-prediction maximal BA value less than -9.54 kcal/mol. These four proteins are all in the opioid receptor family and are highly similar in their 3D structure conformation or 2D sequences. This is validated by the alignments of 3D structures and 2D sequences as shown in Figure S2 of the Supporting information. The heatmap shows more examples of mutual side effects among proteins in one family such as the family of tyrosine kinase protein (JAK1, JAK2, and JAK3), melanocortin receptor (MCR1, MCR3, MCR4, and MCR5), and matrix metalloproteinases (MMP1, MMP2, MMP7, MMP8, and MMP9).

		\subsubsection{Predictions of side effects and repurposing potentials}
		
		The cross-target prediction is a useful tool to detect side effects and to evaluate repurposing potentials of inhibitors. Side effects are caused when a drug candidate exhibits a strong binding affinity to the desired target, but unintentionally acts as a potent inhibitor on other proteins. Drug candidates that exhibit a weak binding affinity to their designated targets but an effective inhibitor to other proteins are deemed to have repurposing potential. Figures 2a and 2b exemplify side effects and repurposing. Each panel involves one target and two off-target proteins. The title, the $x$-axis and the $y$-axis of each panel stand for the target, an off-target protein, and another off-target protein, respectively. The colors of the scattered points indicate the experimental BAs of the inhibitors for the target protein. The red and green colors reveal high and low binding affinities, respectively. The $x$-axis and $y$-axis indicate the predicted BAs from two machine learning models built on inhibitor datasets for two off-target proteins. 
		
\begin{figure}[ht]
			\centering
			\includegraphics[width=0.95\linewidth]{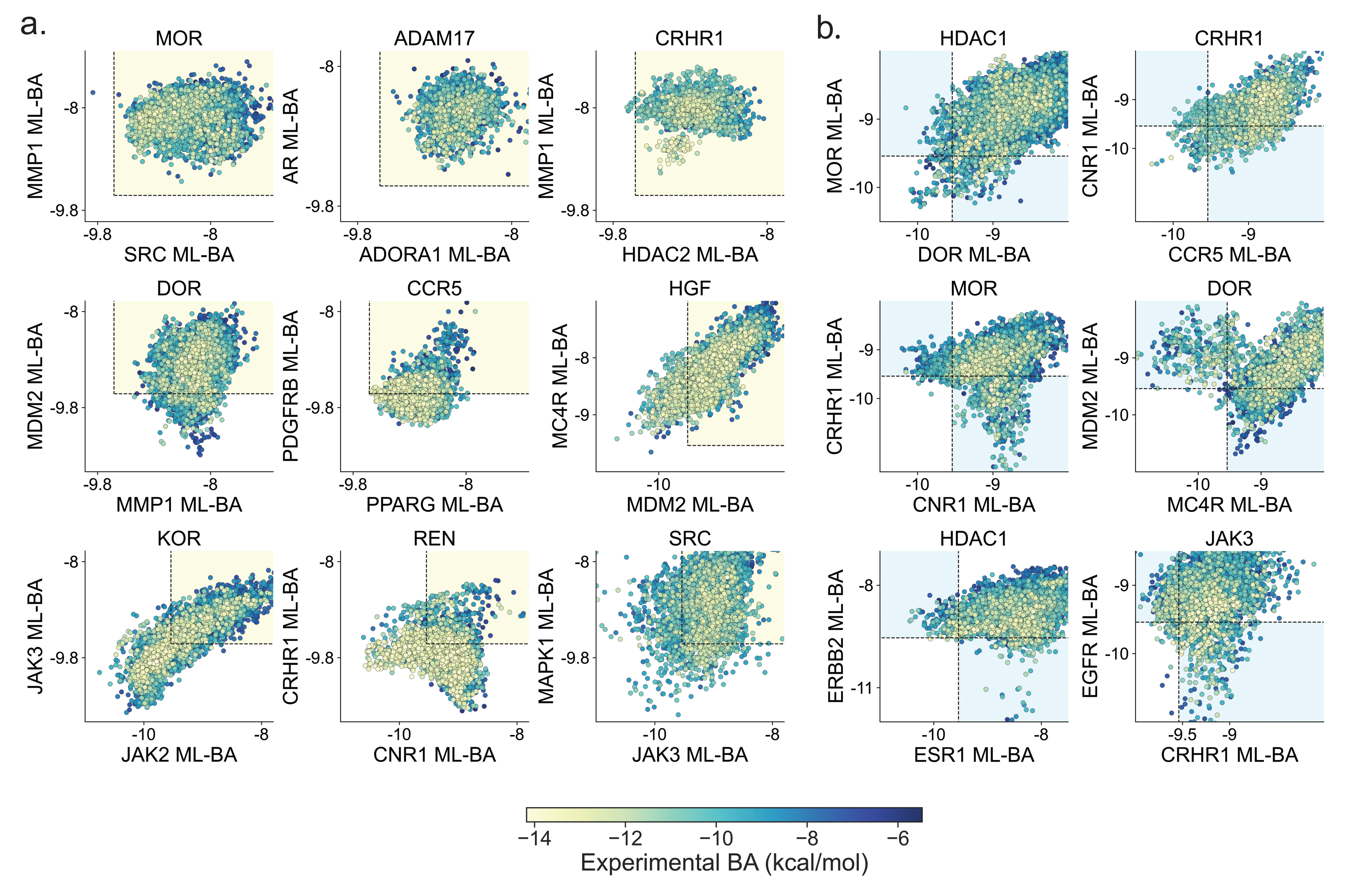} 
			\caption{{\footnotesize 	Examples of inhibitors' predicted side effects and repurposing potentials. The three rows in panel a indicate example inhibitor datasets have side effects on 0, 1, and 2 of the given two off-target proteins, respectively. The yellow frame outlines therein. Yellow zones indicate where side effects are not found. The three rows in panel b reveal example inhibitor datasets that show repurposing potentials on 0, 1, and 2 of the two given off-target proteins. The blue frames highlight the domains where inhibitors have repurposing potential for one protein but have no side effect on the other proteins.} }
			\label{Fig:examples-side-repurpose}
		\end{figure}

		The yellow frames in the nine panels of Figure \ref{Fig:examples-side-repurpose}a highlight the zone where no side effects are induced on two off-target proteins according to our predictions. The three rows in Figure \ref{Fig:examples-side-repurpose}a show some examples of inhibitors for one designated protein having side effects on zero, one, and two of the given two off-target proteins, respectively. For instance, as shown on the second panel in the first row of Figure \ref{Fig:examples-side-repurpose}a, all inhibitors for protein ADAM17 are predicted to be weak inhibitors, i.e., BA values greater than -9.54 kcal/mol, on two off-target proteins. The first panel in the second row shows that around half of the inhibitors for the DOR are predicted to be potent at the MDM2 protein, but all the inhibitors were predicted to not bind to the MMP protein. In addition, the first panel in the third-row exhibits a significant amount of KOR inhibitors that were predicted to be potent at JAK2 and JAK3, simultaneously. 
		
	The repurposing potentials of inhibitors can be revealed through cross-target predictions as well. Figure \ref{Fig:examples-side-repurpose}b provides a few prediction examples of repurposing using our models. The blue frames highlight the zone where inhibitors for target proteins can bind strongly to one protein, i.e., predicted BAs less than -9.54 kcal/mol, but are weaker binders to the other protein, i.e., predicted BAs greater than -9.54 kcal/mol. The first panel in the first row in \ref{Fig:examples-side-repurpose}b shows that many inactive inhibitors for HDAC1 were predicted to have repurposing potential for either MOR or DOR, but not bind to the other one. Since both MOR and DOR are critical targets of medications in treating OUD \cite{wang2019historical}, finding more drug candidates for these two proteins is desirable. Buprenorphine is an FDA-approved drug that is a partial agonist of MOR and KOR, as well as a weak DOR antagonist. As seen in the HDAC1-DOR-MOR panel on the first row of \ref{Fig:examples-side-repurpose}b, there are some inactivate inhibitor compounds for HDAC1 that are effective inhibitors to MOR and DOR. Our models can be used to find more inhibitors that can bind to both targets as Buprenorphine does. The second and third rows in Figure \ref{Fig:examples-side-repurpose}b demonstrate additional examples of the inhibitors for one given protein having repurposing potentials for two other proteins.
		
		\begin{figure}[ht]
			\centering
			\includegraphics[width=0.9\linewidth]{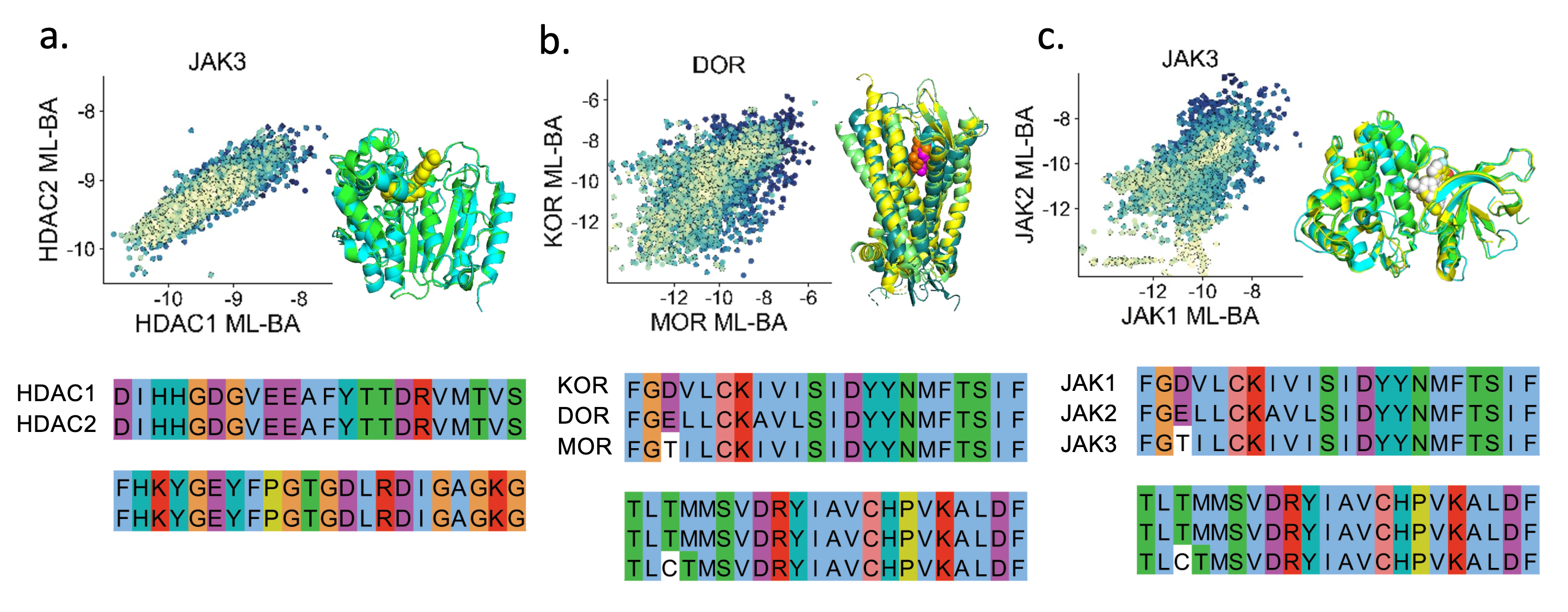} 
			\caption{{\footnotesize Three examples of predicted BA values being correlated. In each example, the chart shows the predicted BA values on two other proteins. The 3D structure alignments are shown on the right of the chart, and the 2D sequence alignment is exhibited on the second row. The 3D structures in the alignment are (PDB 4BKX and 4LY1 for HDAC1 and HDAC2), (PDB 5C1M, 4DJH, 4N6H for MOR, KOR, and DOR), and (PDB 6BBU, 2B7A, and 1YVJ for JAK1, JAK2, and JAK3). } }
			\label{Fig:examples-correlations}
		\end{figure}
		
		\subsubsection{Protein similarity inferred by cross-target BA correlations}

		As discussed above, cross-target BA prediction is useful in evaluating side effects and repurposing potentials. Side effects can be caused when the drug candidate binds to proteins with similar 3D structures or sequences. In such situation, the predicted BA values can be correlated. On the other hand, the correlated predicted BAs can be an indicator of similar binding sites or 3D protein structures. As shown in Figure \ref{Fig:examples-correlations}a, the predicted BAs of inhibitors for JAK3 on HDAC1 and HDAC2 proteins have a nearly linear correlation. The Pearson correlation coefficient (R) of the predicted BA is up to 0.838. This is due to the binding site similarity, which is validated by the 3D protein structure and 2D sequence alignments as shown in \ref{Fig:examples-correlations}a. The 3D structures of the HDAC1 and HDAC2 proteins are found to be quite similar while the 2D sequence identity near the binding site is around 85$\%$. Two more examples of BA correlations revealing similar 3D protein structures are seen in Figures \ref{Fig:examples-correlations}b and \ref{Fig:examples-correlations}c. For the case in Figure \ref{Fig:examples-correlations}b, the Pearson correlation coefficient of the predicted BAs for DOR inhibitors on MOR and KOR proteins is 0.569. For the case in Figure \ref{Fig:examples-correlations}c, the R value of predicted BAs of JAK3 inhibitors on JAK1 and JAK2 proteins is 0.561. The 3D protein structure and 2D sequence alignments confirm the usefulness of cross-prediction in detecting protein similarities. In addition, it was found that there is a bilinear correlation relationship among the predicted BAs and experimental BAs in the case of Figures \ref{Fig:examples-correlations}b and \ref{Fig:examples-correlations}c. The target and two off-target proteins are in the same protein family and share high 3D structure and 2D sequence similarities. A potent DOR inhibitor is likely to be a strong binder on KOR and MOR proteins simultaneously. The high structural similarities form the basis of drug-mediated trilinear target relationship. KOR, MOR, and DOR proteins are often pharmacological targets in the treatment of opioid addiction \cite{veilleux2010review}. The observed bilinear or trilinear relationship indicates the possibility of developing inhibitors that simultaneously bind to multiple targets of the major opioid receptors, namely, MOR, KOR, DOR, and NOR. Such binding effects on multiple opioid receptors have been observed on the currently FDA-approved medications \cite{wang2019historical}.  More examples of similar proteins with correlated predicted BAs can be found in the Supporting information.
		
		\subsubsection{Repurposing to opioid receptors and side effect on hERG}
 		
		MOR, KOR, DOR, and NOR are the four major subtypes of opioid receptors and are the critical pharmacological targets in treating OUD. Inhibitors that bind to these receptors can be potential medications for OUD treatments. We adopted our cross-target prediction  strategy to evaluate the repurposing potential of inhibitors on the four opioid receptors. The 75 collected inhibitor datasets contain more than 120,000  compounds, providing a source of drug candidates in our repurposing study.
		
	 	The side effect of hERG is a priority concern for novel medications, and hence we used our machine learning model to predict the binding affinity of these inhibitors on hERG. A stricter side effect threshold of $-8.18$ kcal/mol  (K$_i=1$ $\mu$M ) was adopted for the hERG. In this study, inhibitors are considered to have no hERG side effect if the predicted BA value on hERG is greater than $-8.18$ kcal/mol. Figures S5 and S6 provide the predicted BAs of the other 73  inhibitor datasets on MOR and hERG. The orange frames highlight the zones where compounds can have repurposing potentials for MOR but do not cause hERG side effects. Some of the 73 inhibitor datasets have almost no compounds in the orange frames such as CXCR2, ERBB4, MAP3K5, PI4KB, and SMO, while other datasets still have a significant number of compounds lying in these orange frames, such as HDAC1, MMP1, MMP2, DOR, KOR, and NOR. 
		
		\subsection{Druggable property screening}
		
		\begin{figure}[ht]
			\centering
			\includegraphics[width=0.85\linewidth]{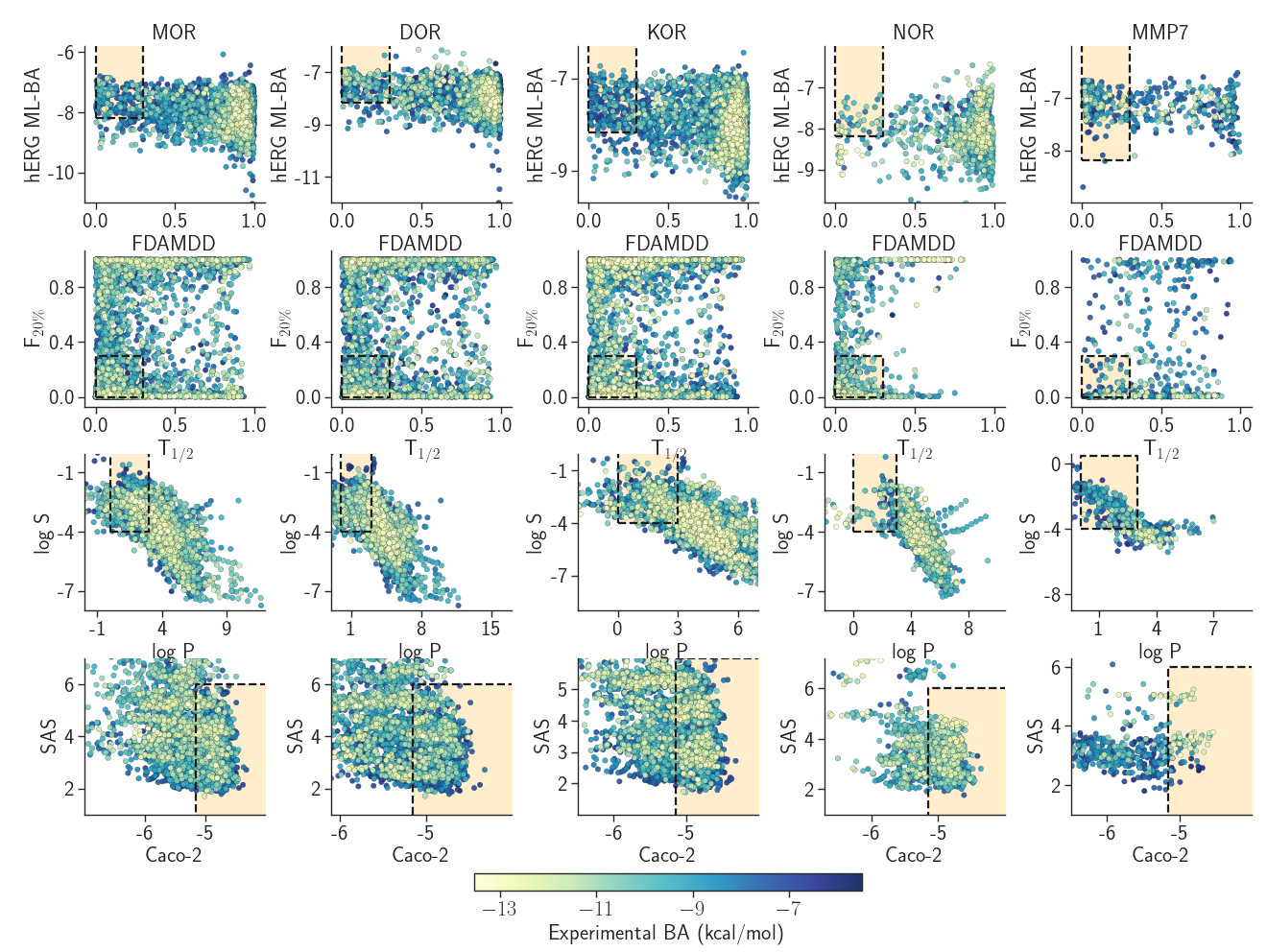} 
			\caption{{\footnotesize Screening of example datasets on ADMET properties, synthesizability, and hERG side effects. The colors of scattered points represent the experimental BA values of inhibitors in each dataset. The Orange frames highlight the optimal ranges of the properties and side effects.} }
			\label{Fig:examples-ADMET}
		\end{figure}
		
		\begin{table}
			\centering
			\begin{tabular}{c|c}		
				\hline
				Property & Optimal range   \\ 	
				\hline
				FDAMDD & Excellent: 0-0.3; medium: 0.3-0.7; poor: 0.7-1.0  \\
				$\rm F_{20\%}$ & Excellent: 0-0.3; medium: 0.3-0.7; poor: 0.7-1.0  \\
				Log P & The proper range: 0-3 log mol/L \\
				Log S & The proper range: -4-0.5 log mol/L \\
				$\rm T_{1/2}$ & Excellent: 0-0.3; medium: 0.3-0.7; poor: 0.7-1.0  \\
				Caco-2 & The proper range: $>$-5.15 \\
				SAS & The proper range: $<$6 \\
				\hline
			\end{tabular}
			\caption{The optimal ranges of six selected ADMET properties and synthesizability (SAS) used to screen nearly optimal compounds.}
			\label{tab:property-optimal}
		\end{table}

	 	ADMET (absorption, distribution, metabolism, excretion, and toxicity) plays a key role in drug discovery and development. It includes vast attributes associated with the pharmacokinetic studies of a compound. A promising drug candidate should not only have sufficient efficacy on the therapeutic target but also satisfies appropriate ADMET properties. Accurate predictions of ADMET are significant in drug design. The successful ADMET screening at the design stage of new compounds is beneficial in reducing the risk of late-stage attrition. To search for promising compounds in treating OUD, systematic screenings of ADMET properties, synthetic accessibility (SAS), and the hERG risk of all inhibitor datasets are needed. We paid attention to six indexes of ADMET, i.e., FDAMDD, T$_{1/2}$ and F$_{20\%}$, log P, log S, and Caco-2, and SAS as well as hERG risk assessment. To evaluate the ADMET properties, we utilized the ADMETlab 2.0 (https://admetmesh.scbdd.com/) solvers that provide machine-learning predictions \cite{xiong2021admetlab,lei2016admet}. Their documents provide a set of optimal ranges of these ADMET properties. The SAS evaluation was obtained from Rdkit packages \cite{landrum2013rdkit}. The optimal ranges of ADMET properties and SAS are provided in Table \ref{tab:property-optimal}, while the BA value $>$ -8.18 kcal/mol is applied as the required range for exempting hERG side effects. With the evaluations of ADMET properties and SAS as well as cross-target prediction tools, we were able to systematically search for promising compound leads. Figure \ref{Fig:examples-ADMET} shows the ADMET screening of a few inhibitor datasets including MOR, DOR, KOR, NOR,  and MMP7. The four rows represent the eight property screenings of the five inhibitor datasets. The colors of the scattered points indicate the experimental BA values of inhibitor compounds.

		FDAMDD is the FDA maximum recommended daily dose, aimed at avoiding toxicity in the human body. The half-life is the amount of time for a drug’s active substance to reduce by half in the human body. The value of T$_{1/2}$ stands for the probability of half-life less than 3 hours. F$_{20\%}$ is the probability of administered drug reaching systemic circulation with less than 20\% of the initial dose. The values of property $\log \rm P$ and $\log \rm S$ are the logarithm of the n-octanol/water distribution coefficient and aqueous solubility value, respectively. Caco-2 is a measure used to estimate in vivo permeability of oral drugs. SAS quantifies the synthesis difficulty of druglike molecules. As seen in Figure \ref{Fig:examples-ADMET}, the orange frames in the panels outline the optimal ranges for a pair of screening properties denoted on the $x$- and $y$-axes. Each pair of screening forms a screening filter for the inhibitors. T$_{1/2}$ and F$_{20\%}$ especially offer a stricter screening as only small portions of inhibitors are covered in the orange frames. The SAS screening seems to be a loose filter, as a significant portion of inhibitors remains in the orange frames after screening. Overall, these ADMET indexes and SAS cause strict restrictions of finding inhibitors.

		\section{Discussion}
		
		\subsection{Side-effect evaluations of existing medications for OUD treatment}

		Substantial pharmacological efforts have been dedicated to the treatment of OUD.  Opioid replacement therapy (ORT) is a popular method to treat people with opioid use disorder. It involves replacing an opioid with a longer-acting but less euphoric opioid. Three classes of medications acting directly on the opioid receptors were found to be effective, namely full agonist, partial agonist, and antagonist. 
		
Methadone, buprenorphine, and naltrexone are approved by the U.S. Food and Drug Administration (FDA) for medication-assisted treatment (MAT). These medications are useful in reducing the risk of death and preventing relapse. In addition, naloxone is a frequently used medication in reducing the risk of overdose, and take-home naloxone is crucial in stopping opioid overdose. 
 It is necessary to evaluate the side effects of these anti-OUD medications, including their actions on five opioid receptors and their multitude of physiological functions affecting the human body.  We used our machine learning models to predict the BA values on the proteins in the five opioid networks as well as on the hERG channel.

			Among the five important opioid receptors, MOR is typically the target of most clinically prescribed medications. Methadone is a full MOR opioid agonist, while it has some agonist effect on KOR and possibly DOR agonist \cite{joseph2000methadone}.  The methadone maintenance treatment (MMT) is beneficial in reducing the intensity of withdrawal symptoms including muscle aches and osteodynia in addicted individuals \cite{joseph2000methadone}. Methadone has a long half-life that makes it more useful in reducing withdrawal symptoms in patients \cite{koehl2019medications}, and consequently reducing patients' compulsive drug-seeking and craving behavior. Our BA predictions of methadone on MOR, KOR, and DOR are -11.8 kcal/mol, -8.96 kcal/mol, and -8.52 kcal/mol, respectively, which agrees with the methadone binding activity on opioid receptors, especially for MOR.  It was reported that methadone prolongs the QT interval in a dose-dependent manner, and high-dose methadone is associated with ventricular tachycardia torsade de pointes \cite{krantz2002torsade}. The overall hERG side effect profile is safe. The predicted BA value on hERG from our model is -7.73 kcal/mol, which is higher than the hERG side-effect threshold of -8.18 kcal/mol, and confirms the safety profile of methadone on hERG. Our predictions indicate that the SMO protein is the only target it can have side effects on. The predicted BA of methadone on SMO protein is -9.67 kcal/mol, and the predicted BAs on all other targets are greater than -9.54 kcal/mol.  SMO is targeted and inhibited by small-molecule drugs for the treatment of advanced basal cell cancer. No serious side effects were reported by inhibiting the SMO protein. Its low side effect profile might be one of the reasons that methadone is the most used medication in MAT and the gold standard against which other medications are compared \cite{bell2020medication}.

		Buprenorphine is a partial agonist of MOR, the antagonist of KOR, and a weak antagonist of DOR \cite{heel1979buprenorphine}. Unlike methadone and other full opioid receptor agonists, buprenorphine has a lower risk of respiratory depression due to a low ceiling to the euphoric effect \cite{walsh1994clinical}. In treating opioid dependence, it is typically administered sublingually as it has an extended half-life than that of intravenous buprenorphine \cite{wang2019historical}, increasing the potential of misuse or overdose. To avoid buprenorphine abuse, it is commonly used with opioid antagonist naloxone via injection or insufflation without causing impairment when used appropriately \cite{bruneau2018management}. The predicted BA values of buprenorphine for MOR, KOR, DOR, and NOR are -12.5, -12.88, -11.64, and -9.41 kcal/mol, which are consistent with the experimental BA values of -12.0, -11.2, -11.7, -9.69 kcal/mol for MOR, KOR, DOR, and NOR, respectively \cite{khanna2015buprenorphine,khroyan2009nociceptin}. Buprenorphine was predicted to have no side effects on hERG with the predicted BA values of -7.31 kcal/mol. It was found to have a minimal impact on the corrected QT interval \cite{wedam2007qt}, which is consistent with our hERG side effect prediction. However, it is predicted to be a potent inhibitor on quite a few other proteins including REN, JAK1, BDKRB1, and NTRK1 with BA values of -10.74,  -10.67, -10.49, and -10.28 kcal/mol. JAK1 is a member of the Janus kinase family and Janus kinase inhibitors are used in the treatment of cancer and inflammatory diseases. Clinical trials indicated inhibitors for NTRK1 protein have shown efficacy as targeted therapies for extracranial tumors. The inhibition of buprenorphine needs to be further investigated for its side effects on previously discussed proteins and its usefulness in the treatment of alternative diseases. 
		
		LAAM, acting as a MOR agonist, can provide greater suppression of heroin use in comparison to methadone. Our predicted BAs of LAAM to MOR, DOR, KOR, and NOR are -9.34, -8.94, -8.83, and -9.69 kcal/mol, respectively. The potency of LAAM on these receptors is not as strong as methadone and buprenorphine. In addition, the safety profile of LAAM is low due to its potential for ventricular rhythm disorders \cite{deamer2001torsades}. Such adverse effects are directly associated with the hERG blockade, which is verified by the predicted relatively high BA of -7.94 kcal/mol on hERG, a value close to our hERG side effect threshold. According to our models, the top potential targets with side effects imposed are SMO and JAK1 proteins with predicted BA values of -9.82 and -9.75 kcal/mol. Similar to the predictions for buprenorphine, the molecular binding on SMO and JAK1 might not cause a serious problem. 
		
		Naltrexone is an antagonist of MOR and a partial agonist of KOR. Long-acting injectable naltrexone can block opioid receptors but does not activate them, reducing drug-seeking behavior and alleviating drug craving  \cite{lapham2011open}. Naltrexone is observed to have a continuous effect in reducing the frequency and dosage of heroin use \cite{skolnick2018opioid} and in decreasing the risk of opioid overdose. Naltrexone and nalmefene have a longer duration period, therefore drawing research and clinical interests to investigate their anti-overdose effect against potent fentanyl analogs \cite{krieter2019fighting}. We predicted BA values for MOR, KOR, NOR, and DOR respectively -12.54, -12.05, -9.88, and -10.49 kcal/mol. The predicted BA value on hERG was low with a value of -7.64 kcal/mol, suggesting a low hERG side effect potential. Strong binding potency can occur on a few proteins including JAK1, BDKRBA, and SMO with the predicted BA values of -10.30, -10.02, and -9.92 kcal/mol, respectively. Analogous to naltrexone, nalmefene is also a MOR antagonist and a KOR partial agonist. It has a prolonged duration of action and intravenous doses of nalmefene have been shown effective at counteracting the respiratory depression produced by an opioid overdose \cite{park2019prevention}. It was predicted to be potent at the four opioid receptors MOR, KOR, DOR, and NOR with BA values of -12.62, -12.11, -10.78, and -10.05 kcal/mol. The proteins it can bind to, with a strong binding affinity, include JAK1, SMO, BDKRB1, REN, and S1PR1 with predicted BA values of -10.25, -10.08, -10.06, -9.99, and -9.99 kcal/mol. Protein BDKRB1 is a G-protein coupled receptor that mediates responses to pathophysiologic conditions such as inflammation, trauma, burns, shock, and allergy. Antagonist inhibitors of this receptor were used to reverse acute or persistent inflammatory pain in these pathophysiologic conditions. Naltrexone or nalmefene might have some clinical significance as pain relief in these pathophysiologic conditions. Activation of receptor protein S1PR1 is heavily involved in immune cell regulation and development. It is also responsible for vascular growth and development, during embryogenesis. Inhibitions of protein S1PR1 by naltrexone may interfere with normal growth and development.

		Naloxone is a non-selective and competitive opioid antagonist that reverses opioid analgesic actions quite effectively. Naloxone is commonly used for acute opioid intoxication, rescuing opioid-induced respiratory depression \cite{park2019prevention}. However, it is of lower potency and shorter duration period, compared to other antagonists such as naltrexone and nalmefene. In addition, the capabilities of naloxone are limited when ingesting highly potent opioids, such as fentanyl \cite{national2017naloxone}. The predicted BA value on MOR, KOR, and DOR are -11.50, -10.93, and -9.79 kcal/mol, close to those experimental BA values of -12.21, -10,63, -10.75 kcal/mol, respectively. The BA prediction for hERG is -7.58 kcal/mol, suggesting a safe hERG-blockade profile. The potential side effect on other proteins included JAK1, JAK3, REN, BDKRB1, PDGFRB, and ERBB4 with predicted BA values of -10.34, -9.92, -9.89, -9.85, -9.80, and -9.76 kcal/mol.  Like naltrexone and nalmefene, naloxone could also block protein BDKRB1, and may also be clinically useful in the pain relief of several pathophysiologic conditions. Protein PDGFRB is essential for vascular development, and its inhibition may compromise the integrity and/or functionality of the vasculature in multiple organs, including the brain, heart, kidney, skin, and eyes \cite{soriano1994abnormal}. HER4 is a receptor tyrosine kinase that is critical for normal body systems such as the cardiovascular, nervous, and endocrine systems. Overexpression of HER4 kinase results in cancer development \cite{el2021review}, and activation of HER4 by ligand binding can potentially cause cancer, promoting drug development to inhibit these HER4\cite{el2021review}. The duality of naloxone’s ability to activate HER4, potentially causing cancer or providing an effective HER4 inhibitor, needs to be investigated further.
	
		Heroin (Diamorphine), a MOR agonist, was found to be useful in helping patients disengage from the use of street heroin and reducing criminal involvement. It can be an effective adjunctive treatment for chronic, relapsing opioid dependence \cite{oviedo2009diacetylmorphine}. Heroin assisted treatment is now  available in Canada and some European countries as a new treatment modality. It is administrated under direct medical or nurse supervision. Heroin assisted treatment is intended for injection into patients suffering from OUD, who have not responded to standard medications for OUD. However, compared to other medications for opioid use disorder, its safety profile is low with major adverse effects, such as respiratory depression and seizures \cite{oviedo2009diacetylmorphine}. The predicted BA of heroin on MOR by our model is -9.46 kcal/mol. In addition, it was predicted to be a potent inhibitor at NOR, JAK1, JAK3, BDKRB1, SMO, REN, and ITGB1 with BA values of -10.67, -10.32, -10.15, -10.10, -9.87, -9.77, and -9.72 kcal/mol. The BA value on hERG was predicted to be -7.59 kcal/mol. Protein  ITGB1   associates with integrin alpha 1 and integrin alpha 2 to form integrin complexes which function as collagen receptors. Recent studies have shown that the inhibition of ITGB1 enhances the anti-tumor effect of cetuximab in colorectal cancer cell \cite{yang2020inhibition}. Nonetheless, due to the severe addiction effect, heroin may not be a good choice for cancer treatment.

		Hydromorphone is also a MOR agonist. Injectable hydromorphone was also found to be as effective as diacetylmorphine for patients who have not benefited from previous treatments, such as methadone or suboxone \cite{bansback2018cost}. Following studies showed that once-daily sustained-release oral hydromorphone was useful in managing cravings without notable side effects. It has the advantage of no influence on the cardiac QTc interval \cite{braithwaite2020sustained}. The treatment with hydromorphone requires supervised administration. In our prediction, hydromorphone had strong binding affinities on all four opioid receptors, namely, MOR, KOR, NOR, and DOR with the BA values of -12.9, -11.58, -10.35, and -10.08 kcal/mol, respectively. The predicted BA value on hERG is -7.71 kcal/mol, which can be deemed as having a low potential of side effect on hERG. Other targets that hydromorphone can possibly cause side effects on are JAK1, AR, PDGFRB proteins with the predicted BA values of -10.20, -9.81, and -9.78 kcal/mol. 
		
		Dihydrocodeine is a semi-synthetic opioid analgesic and agonist. It is sometimes used for maintenance treatment as an alternative to methadone or buprenorphine in some European countries. Our predicted BA values of dihydrocodeine on MOR, DOR, KOR, and NOR are -9.41, -7.44, -7.91, and -11.06 kcal/mol. It may be used as a second line treatment. Low quality evidence reported that dihydrocodeine may be no more effective than other routinely used medication interventions in reducing illicit opiate use \cite{carney2020dihydrocodeine}. It can be a potent inhibitor for several proteins such as JAK1, SMO, JAK3, BDKRB1, ITGB1, and AR, with predicted BA values of -10.07, -9.89, -9.83, -9.76, -9.73, and -9.68 kcal/mol. The predicted BA value on hERG was -8.08 kcal/mol, which shows a mild potential of side effects on hERG and proteins ITGB1 and AR. Androgen receptor (AR) functions mainly as a DNA-binding transcription factor that regulates gene expression. High expression in androgen receptor has been linked to aggression and sex drive \cite{cunningham2012androgen}. AR also has roles in the progression of prostate cancer and is an important therapeutic target in prostate cancer \cite{helsen2014androgen}. The inhibition of AR by Dihydrocodeine may have an impact on male sexual phenotype.

			Lofexidine is an $\alpha_2$-adrenergic receptor agonist but is not classified as an opioid. It is an alternative for people with mild or uncertain opioid dependence in need of short-term detoxification and is effective in reducing withdrawal symptoms of OUD. Its adverse side effects include QT prolongation. Our predicted BA values for MOR, KOR, DOR, and NOR are -8.33, -8.53, -8.02, and -9.7 kcal/mol, which are consistent with the fact that Lofexidine is not an opioid. Our BA prediction of Lofexidine on hERG is -7.30 kcal/mol, showing a good side profile. The strongest predicted BAs were for FDE4A, SMO, and IITGB1 with the values of  -9.79, -9.76, and -9.56 kcal/mol. The side effects of these three proteins might be mild.

		Currently, there are some drugs used in the treatment of opioid-induced constipation such as alvimopan, methylnaltrexone, and naloxegol. The three drugs are all peripherally acting $\mu$-opioid receptor antagonists due to their limited ability to cross the blood–brain barrier and reach the MORs of the central nervous system. Consequently, the effects of centrally-acting opioid antagonists are normally not notable. Alvimopan is approved for the treatment of postoperative due to its intended blockade of MORs in the gastrointestinal tract. We predicted BA values for the MOR, KOR, NOR, and DOR  -12.40, -9.50, -10.18, and -11.02 kcal/mol, respectively. Its potency on MOR is validated in our prediction. It can also be potent at several proteins including JAK1, BDKRB1, S1PR1, ACE, and SMO with predicted BA values of -10.54, -10.36, -10.28, -10.07, and -10.07 kcal/mol. The most common side effects of alvimopan include dyspepsia, hypokalemia, back pain, and delayed micturition. Methylnaltrexone is used to treat opioid-induced constipation in chronic non-cancer pain or when ordinary laxatives do not work well. It was also predicted to be a potent inhibitor of MOR, KOR, and NOR with BA values of -11.04, -10.62, and -10.25 kcal/mol. Potential side effects can be exhibited on proteins including JAK1, BDKRB1, and REN with BA values of -10.57, -10.05, and -10.04 kcal/mol, respectively. Naloxegol is recommended for the treatment of opioid-induced constipation (OIC) in patients with chronic non-cancer pain. Its predicted BA values for NOR, KOR, MOR, and DOR are -10.52, -10.15, -9.96, and -9.26 kcal/mol, respectively. Based on our predictions,  it can be potent for    REN, JAK1, BDKRB1, and ERBB4 with predicted BA values of -10.86, -10.59, -10.50, and -10.03 kcal/mol, respectively.
	
		\subsection{Nearly optimal lead compounds from screening and repurposing}
		
		As previously discussed, opioid replacement therapy (ORT) replaces an opioid with a longer-acting but less euphoric opioid, and is widely used in OUD treatment. Drugs like methadone and buprenorphine are commonly used for ORT, and act as agonist/antagonist depending on the different opioid receptors. We dedicate our efforts to search for more potential inhibitors of the four opioid receptors, which can be potential agonist/antagonist in OUD treatments. Screening and repurposing are the two avenues to find more inhibitors. There were 74 models utilized to predict the cross-target bind affinity in the process of screening and repurposing. In addition to the potency concern, optimal ranges for the ADMET properties, synthetic accessibility in Table \ref{tab:property-optimal}, and hERG side effect all needed to be satisfied. As we know, MOR, DOR, KOR, and NOR are the four major opioid receptors, and critical pharmacological targets in OUD treatments. In finding more promising potent compounds at these four receptors, the 74 inhibitor datasets are our source of inhibitor compounds. In the screening process, we start with potent inhibitor compounds (experimental BA values $<$ -9.54 kcal/mol) in the inhibitor datasets of the four opioid receptors, and then evaluate a series of other properties. It is important to note that if a designated inhibitor of a receptor exhibits notable potency on any of the other three receptors, it is not considered as a side effect. It is very common that one inhibitor can be effective on several of the four major opioid receptors simultaneously, as seen by the currently approved drugs that act as agonists or antagonists on several receptors. However, the potential for side effect concern must be evaluated on the other 70 protein targets including hERG. We require the predicted BA values $>$ -9.54 kcal/mol to exempt side effects except for hERG with a stricter BA $>$ -8.18 kcal/mol. In the repurposing process, we evaluate the binding potency of all weak inhibitors in the other 70 datasets on the four opioid receptors. We start with inhibitors of experimental BA value $>$ -9.54 kcal/mol, and then find those with predicted BA values $<$ -9.54 kcal/mol on the four opioid receptors. In search of inhibitors with repurposing potential on the opioid receptors, their side effects need to be exempted on the other 70 proteins. Following these, optimal range of ADMET properties and synthetic accessibility need to be screened.

\begin{figure}[ht]
			\centering
			\includegraphics[width=0.9\linewidth]{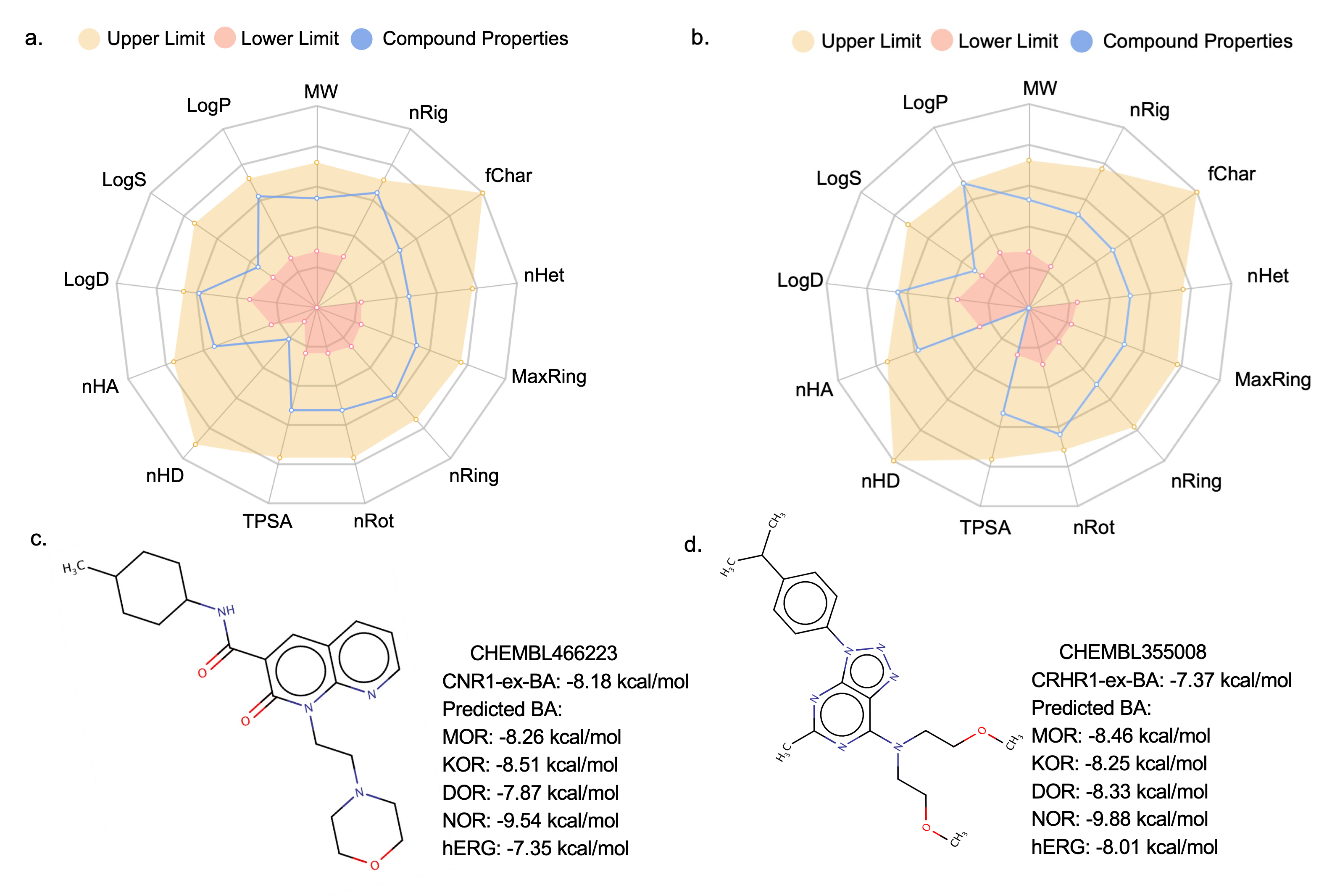} 
			\caption{{\footnotesize Evaluations of more ADMET properties for the found molecular compounds with repurposing potentials. Panel a and c represent the prediction results of ADMET properties and side effect evaluations for compound ChEMBL466223, and panels b and d stand for these predictions for compound ChEMBL355008. The boundaries of yellow and red zones in figure a and b highlight the upper and lower limits of the optimal ranges for the ADMET properties, respectively. The blue curves represent values of the specified 13 ADMET properties. Figures a and b are the prediction results from ADMETlab 2.0 (https://admetmesh.scbdd.com/) website. 
Abbreviations: 
MW (Molecular Weight), 
$\log$P (log of octanol/water partition coefficient),  
$\log$S (log of the aqueous  solubility), 
$\log$D (logP at physiological pH 7.4), 
nHA (Number of hydrogen bond acceptors), 
nHD (Number of hydrogen bond donors), 
TPSA (Topological polar surface area), 
nRot (Number of rotatable bonds), 
nRing (Number of rings), 
MaxRing (Number of atoms in the biggest ring), 
nHet (Number of heteroatoms), 
fChar (Formal charge), and 
nRig (Number of rigid bonds).} }
			\label{Fig:repurpose-ADMET}
		\end{figure}

	Two inhibitor compounds, CHEMBL466223 from the CNR1 dataset and CHEMBL355008 from the CRHR1 dataset,   were found to satisfy all the aforementioned criteria for repurposing. They were predicted to be effective on NOR with predicted BA values of -9.543 kcal/mol and -9.88 kcal/mol while their experimental BA values for the designated targets are -8.18 kcal/mol and -7.37 kcal/mol. They were not predicted to be potent at the other three opioid receptors, MOR, DOR, and KOR, as seen by the following predicted BA values of -8.26, -7.87, and -8.51 kcal/mol, respectively. Its predicted BA value on hERG was -7.34 kcal/mol. The predicted BA values of compound CHEMBL355008 on MOR, DOR, and KOR were -8.46, -8.33, and -8.25 kcal/mol, respectively, and -8.01 kcal/mol on hERG. These two compounds were predicted to have no binding effects or side effects on all other 69 proteins. We carried out evaluations of more ADMET properties regarding these two molecular compounds using the ADMETlab 2.0 predictive solver. As seen in Figures \ref{Fig:repurpose-ADMET}a and \ref{Fig:repurpose-ADMET}b, the two compounds are still in the optimal ranges of these ADMET properties. The meaning and optimal ranges of the 13 ADMET properties are provided in the Supporting information. We are also interested in the molecular interaction between the two inhibitors and NOR protein structure and utilized the software AutoDock Vina to perform protein-ligand docking to analyze the interactions. The 3D docking structure and 2D interaction diagrams are shown in Figure \ref{Fig:ducking-ligplot}. It can be seen in the figure below, hydrogen bonds are formed between the inhibitors and the NOR protein. Compound CHEMBL355008 has two hydrogen bonds formed with the Tyr131 (3.32 $\mathring{\rm A}$) and Tyr309 (3.17 $\mathring{\rm A}$), respectively. The compound CHEMBL466223 has one strong hydrogen bond (2.82 $\mathring{\rm A}$) with Tyr309. The predicted binding energies by CHEMBL355008 and CHEMBL466223 with NOR were -9.54 and -9.88 cal/mol, respectively. The single strong hydrogen bond of CHEMBL466223 partially explains its predicted higher binding energy with NOR even though CHEMBL466223 has two hydrogen bonds with NOR. It is observed that the side chains of Tyr309 play roles in the formation of hydrogen bonds with the two compounds. No covalent bond is formed by either of the two compounds with the side chains of the NOR protein, and hydrogen bonds play the essential roles of binding energy. 
		
		Even though these two compounds were not predicted to be potent inhibitors on MOR, similar to the approved drugs for OUD treatment, they are selective inhibitors for NOR. Selective antagonists are needed in scientific research when one of the receptors needs to be blocked without affecting the others. These two were predicted to be selective inhibitors of NOR with safe side effect profiles on all other 70 proteins. Further studies can be carried out to test their physiological effect on NOR or their pharmacological effect in treating OUD.
		
		\begin{figure}[ht]
			\centering
			\includegraphics[width=1.0\linewidth]{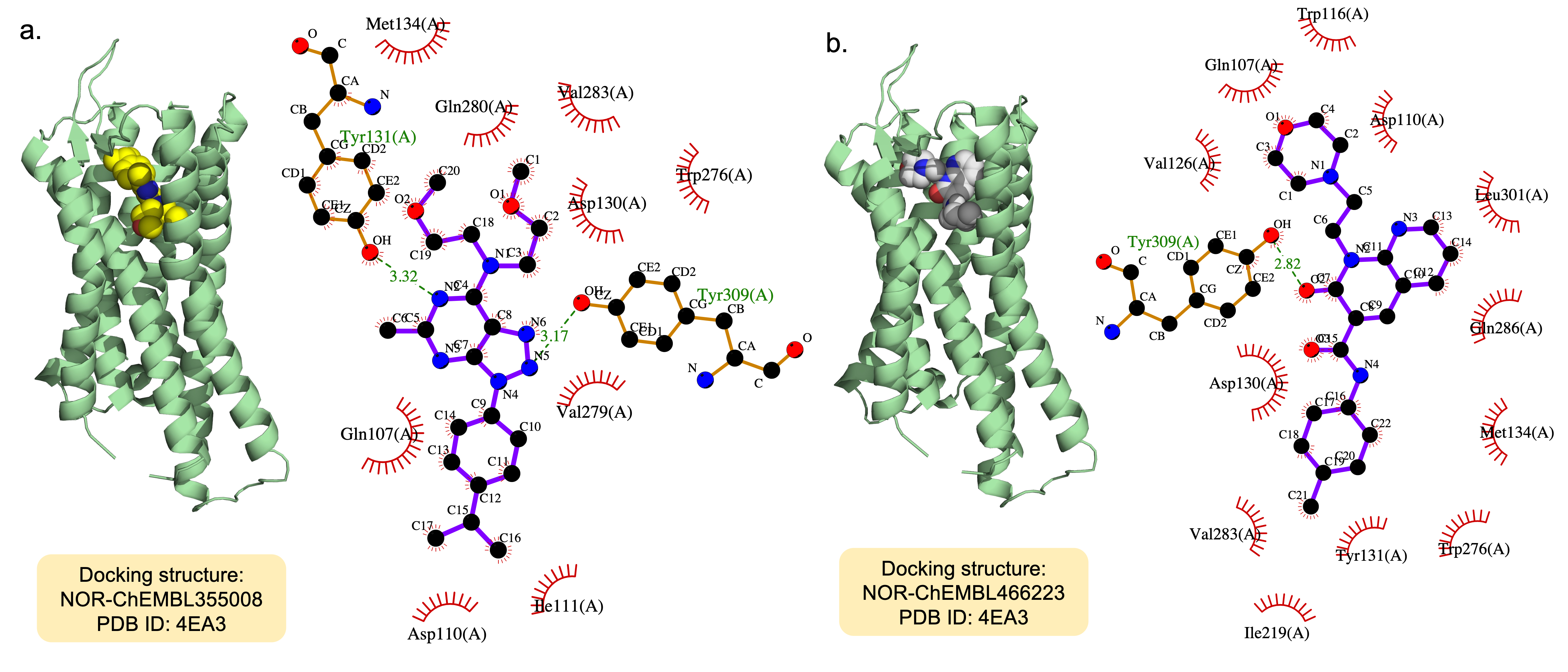} 
			\caption{{\footnotesize 	The docking structure of our nearly optimal lead compounds bound to nociceptin opioid receptor and their 2D interaction diagrams. AutoDock Vina was used to performing the protein-ligand docking. Hydrogen bonds are playing essential roles in binding energies.} }
			\label{Fig:ducking-ligplot}
		\end{figure}

		\section{Methods}

		\subsection{Datasets}
		
		The inhibitor datasets were collected from the ChEMBL database for the proteins in the five investigated opioid receptor networks. As machine-learning models rely on a training set of sufficient data points, we require the size of the collected inhibitor dataset to be at least 250. A total of 74 datasets were then obtained. The labels for the data points are IC$_{50}$ or $\rm K_{i}$. As suggested in \cite{kalliokoski2013comparability}, the IC$_{50}$ values can be approximately converted to K$_i$ values with the relation K$_i$=IC$_{50}/2$. These labels were used to compute the binding affinity (BA) with the formula BA=1.3633$\rm \times \log_{10} K_{i}$ (kcal/mol), which were then used in building machine learning models. Since the hERG is a critical target of side effects in drug design, an inhibitor dataset was also collected for it from the ChEMBL database. The details regarding all these datasets are provided in the Supporting information.

		\subsection{Molecular embeddings}
		
		The molecular representation for the inhibitors in the collected 75 datasets is  2D SMILES strings. Two forms of molecular fingerprints were used to build machine-learning models in this study. The molecular fingerprints were generated by pre-trained models based on natural language processing (NLP) algorithms including transformer \cite{chen2021extracting} and sequence-to-sequence autoencoder \cite{winter2019learning}. The two pre-trained models encode the 2D SMILES strings of inhibitor compounds in latent embedding vectors of lengths 512. We denote the two types of fingerprints by the transformer and autoencoder models as TF-FP and AE-FP, respectively.
		
		\subsubsection{Sequence-to-sequence auto-encoder }

		A data-driven unsupervised learning model was recently proposed to extract molecular information embedded in the SMILES representation \cite{winter2019learning}. A sequence-to-sequence autoencoder was utilized to translate one form of molecular representation to another, with a comprehensive description of the chemical structure compressed in the latent representation between the encoder and decoder. The translation model extracts the physicochemical information in the molecular representation when translated to another semantically equivalent but syntactically different representation of the molecule. 
		The translation model was trained on a large set of chemical structures and allows for the molecular descriptor extraction for query compounds without retraining or using labels. 
		
		The translation model consists of the encoder and decoder networks. The information bottleneck in between is used to compress the essential information of the input SMILES, and the embedded information is then used as input in the translation through the decoder. In the encoder network, convolutional neural network (CNN) and recurrent neural network (RNN) architectures were adopted. Then fully connected layers map the output of CNN or the concatenated cell states of the RNN networks to intermediate vector representations between encoders and decoders. The decoder is comprised of RNN networks with latent vectors as input. To embed more meaningful physicochemical information about molecules in the latent vectors, a classification model was used to extend the translation model by predicting certain molecular properties based on the latent vectors. The output of the decoder's RNN network is the probability distributions over different characters in the translated molecular representations. In training the autoencoder model, the loss function is the sum of cross-entropies between probability distributions and one-hot encoded correct characters as well as the mean squared errors for molecular property predictions from the classification model. 
		
		The translation model was trained with approximately 72 million molecular compounds from ZINC and PubChem databases. The preprocessing was carried out to filter compounds with a variety of criteria including molecular weight, heavy atom numbers, partition coefficient, and other properties. After sufficient training with the processed dataset, the resulting translation model yields the embedding vectors as molecular fingerprints.
		
		\subsubsection{Bidirectional transformer }
		
		A self-supervised learning (SSL)-based platform was recently developed to pre-train a deep learning network from millions of unlabeled molecules. Predictive molecular fingerprints can be extracted from the pre-trained models \cite{chen2021extracting}. The self-supervised learning was achieved with the bidirectional encoder transformer (BET) model that relies on the attention mechanism. The SSL has the advantage of avoiding the construction of a complete encoder-decoder framework and solely using the decoder network to encode the SMILES of molecules.

The SMILES strings of molecules were the input for the SSL pretraining platform. Pairs of real SMILES and masked SMILES were constructed by hiding a certain percentage of some meaning symbols in the strings. Then an SSL approach enables the model training with the data-mask pairs in a supervised way. In the pretraining process, the symbols of masked symbols were referred to by learning the unprocessed ones in SMILES, which then leads to the understanding of SMILES language. Data masking is preprocessed before starting to train the model with SSL. A total of 51 symbols were considered as the components in the SMILES strings. The SMILES were the input for training the model, and we required the maximal length to be 256. Symbols $'\langle s \rangle'$ and  $'\langle \backslash s \rangle'$ were added to the beginning and the end of SMILES strings. If the length is less than 256, the symbol $'\langle pad \rangle'$ was used to supplement a SMILES string. For the data masking, a total of 15$\%$ of the symbols in all the SMILES were operated, among which $80\%$ were masked, $10\%$ were unchanged and the remaining $10\%$ were randomly changed.

		The BET modules play critical roles in achieving SSL from a massive number of SMILES strings. The attention mechanism in transformer modules captures the importance of each symbol in the inputted SMILES sequences. The BET consists of eight bidirectional encoder layers, with each encoder layer composed of a multi-head self-attention layer and a subsequent fully connected feed-forward neural network. The number of heads in each self-attention layer is 8, and the embedding size of fully connected feed-forward layers is 1024. The Adam optimizer was used in the training process and weight decay of 0.1 was applied. The loss function is defined to be the cross-entropy, measuring the difference between the real and predicted symbols at masked positions. The maximum length of input SMILES is 256 including the added special symbols at the two ends, while the embedding dimension of each symbol was 512. The resulting molecular embedding matrix is comprised of 256 embedding vectors of dimension 512. The mean of embedding vectors for the valid symbols in one SMILES string was used as molecular fingerprint of a given SMILES. 
		
         Due to the high parallelism capability and training efficiency from transformer modules, a massive number of SMILES can be used to train deep learning models. In our implementations, SMILES strings from one or the union of the ChEMBL, PubChem, and ZINC databases were employed, giving rise to three pre-trained models \cite{chen2021extracting}. In this study, transformer-based embeddings generated from the pre-trained model solely using the ChEMBL database were used as molecular fingerprints.


		\subsection{Machine-learning models}
		
		The gradient boosting decision tree (GBDT) algorithm was deployed to build our machine learning models. The GBDT algorithm is a popular ensemble method and has the advantage of robustness against overfitting, insensitiveness to hyperparameters, and ease of implementation. The methodology is to create many weak learners (individual trees) by bootstrapping training samples and to make predictions by integrating the outputs of weak learners.  Weak learners are likely to make poor predictions, but through ensemble approach the overall errors by combining all the weaker learners are reduced. GBDT is particularly useful when training with small datasets and can deliver better prediction performance than deep neural network (DNN) and some other machine learning algorithms. It gains wide popularity in a range of quantitative structure–activity relationship (QSAR) prediction problems \cite{cang2017analysis,jiang2020boosting}, and promotes the development of competitive predictive ML models. The GBDT algorithm provided in the Scikit-learn (version 0.24.1) library was used in this work. 
		
		We collected a total of 75 inhibitor datasets with at least 250 data points in each dataset. It is preferable to utilize GBDT in building models for these datasets. As aforementioned, two types of molecular fingerprints including TF-FP and AE-FP were adopted to represent inhibitor compounds. Our machine-learning (ML) models were built by integrating these molecular fingerprints with the GBDT algorithm. We built a total of 75 ligand-based ML models with the 75 inhibitor datasets. For each dataset, two individual models were built by pairing TF-FP and AE-FP with GBDT algorithm, and then the average of the predictions from the two individual models was regarded as our final binding affinity prediction. Such average or consensus results typically outperform those from individual models. To alleviate the effect of randomness, each of the individual GBDT models were trained ten times with a different random seed. The average of the ten predictions was regarded as the final outcome of each individual model. In the Supporting information, we included the Pearson correlation coefficients of five-fold cross validations for modeling the 75 datasets.
		
		\begin{figure}[ht]
			\centering
			\label{Fig:ligplot}
		\end{figure}

		\section{Conclusion}
		
	 	Opioid use disorder (OUD) is a chronic and complex disease with neurobiological, psychological, behavioral, and medical implications. Each year in the United States and around the world, thousands of deaths are caused by opioid abuse, and billions of dollars have been spent on OUD treatment. To combat the opioid epidemic, efforts in novel treatment formulations and devices have been dedicated by pharmaceutical agencies and scientists. Pharmacological or psychosocial interventions showed their efficacy for OUD treatment, but many patients still drop out of treatment and return to opioid-dependent life because of the chronic and relapsing nature of opioid addiction. The development of nonaddictive analgesics and anti-opioid vaccines can be potentially effective in opioid abuse prevention and the OUD treatment, but the progresses seem very slow. More options for treatment are needed to combat such destructive diseases.

		Opioid receptors are the direct targets of opioids, and medications on them are found effective in opioid addictions. OUD affects intricate molecular and biological activities in the brain involving significant protein-protein interactions (PPI) in various brain areas. The development of anti-OUD medications cannot neglect the impact of opioids or medications on the PPI networks of opioid receptors. In this work, we developed proteome-informed machine learning protocol to study OUD and discover more drug candidates to treat it. With molecular fingerprints generated by advanced NLP models based on transformer and autoencoder algorithms, gradient boosting decision tree (GBDT) algorithm was used to build our predictive models. The consensus predictions from two forms of molecular fingerprints could enhance the predictive performance. We used these models to reevaluate the side effects of currently available medications for treating OUD. In addition, these models were used to study the repurposing potentials of existing inhibitors on the major opioid receptors and screened the possible side effects of these inhibitors. The evaluations of ADMET properties were then carried out with machine-learning predictions. We identified a group of promising compounds targeting the opioid receptors. Considering the therapeutic efficacy by antagonist or agonist effect of currently approved drugs, further animal experiments with these compounds are needed to test the antagonist/agonist properties. More tests in vitro or animal arrays are needed to scrutinize the toxicity and blood-brain barrier permeability characteristics of these candidate compounds. Automated generation of more drug candidates can be carried out using our generative network modules \cite{gao2020generative}, and this study can be employed for the screening of potential side effects. 
		
		Our machine-learning-based platform provides a novel approach for searching compound candidates to treat OUD and can be generalized to the studies of other diseases with neurological implications. With more advances in understanding the opioid addiction mechanism and more efforts from pharmacological treatment, our platform can be assistive in combating the serious public health issues from OUD.
		
		\section*{Data and code availability}
		
		The related datasets studied in this work are available at: 
		https://weilab.math.msu.edu/DataLibrary/2D/. Codes are available at https://github.com/WeilabMSU/OUD-PPI.

		\section*{Acknowledgment}
		This work was supported in part by NIH grant  GM126189, NSF Grants DMS-2052983,  DMS-1761320, and IIS-1900473,  NASA 80NSSC21M0023,  MSU Foundation, Michigan Economic Development Corporation,  George Mason University award PD45722,  Bristol-Myers Squibb 65109, and Pfizer.

		%
		%

\end{document}